\documentclass[preprint,tightenlines,aps,prd,showpacs,superscriptaddress,nofootinbib,floatfix]{revtex4}
\usepackage{amsfonts}
\usepackage{multirow}
\usepackage{mathrsfs}
\usepackage{graphicx}
\usepackage{amsmath}
\usepackage{amssymb}
\usepackage{bm}
\usepackage{bbm}
\usepackage{slashbox}
\usepackage{textpos}

\renewcommand{\l}{\left}
\renewcommand{\r}{\right}

\newcommand{\tn}{\textnormal}

\begin{document}
\title{Light quark mass dependence of the $\bm{X(3872)}$ in XEFT}

\author{M. Jansen}
\affiliation{Helmholtz-Institut f\"ur Strahlen- und Kernphysik,
Universit\"at Bonn, 53115 Bonn, Germany}
\affiliation{Institute of High Energy Physics, Chinese Academy of Sciences, 
Beijing 100049, China}

\author{H.-W. Hammer}
\affiliation{Helmholtz-Institut f\"ur Strahlen- und Kernphysik,
Universit\"at Bonn, 53115 Bonn, Germany}
\affiliation{Institut f\"ur Kernphysik, 
Technische Universit\"at Darmstadt, 64289 Darmstadt, Germany}
\affiliation{ExtreMe Matter Institute EMMI, GSI Helmholtzzentrum 
f\"ur Schwerionenforschung GmbH, 64291 Darmstadt, Germany}

\author{Yu Jia}
\affiliation{Institute of High Energy Physics, Chinese Academy of Sciences, 
Beijing 100049, China}
\affiliation{Theoretical Physics Center for Science Facilities, 
Institute of High Energy Physics, Chinese Academy of Sciences, 
Beijing 100049, China}

\date{\today}
\begin{abstract}
The quark mass dependence of hadrons is an important input for lattice
calculations. We investigate the light quark mass dependence of
the binding energy of the $X(3872)$ and the $\bar{D}^0D^{*0}$ scattering 
length in the $C=+1$ channel to next-to-leading order in XEFT where 
pion interactions are perturbative. At this order, the quark mass 
dependence is determined by a quark mass-dependent contact 
interaction in addition to the one-pion exchange. Using naturalness arguments
to constrain unknown parameters, we find a moderate sensitivity of the 
binding energy for quark masses up to twice the 
physical value while the scattering length is more sensitive. The occurrence of
infrared divergences due to on-shell pions in XEFT and their treatment is discussed in detail.
\end{abstract}




\maketitle
\section{Introduction}
The discovery of a flurry of new quarkonium-like hadrons in the last decade
has created exciting prospects in quarkonium physics \cite{Brambilla:2010cs}. 
In 2003, the Belle Collaboration discovered a charmonium-like hadron \cite{Choi:2003ue}, known as the $X(3872)$, which was quickly
confirmed by the CDF collaboration \cite{Acosta:2003zx}. Its observed decays into $J/\psi \gamma$ imply even charge 
parity \cite{Aubert:2006aj}. Ending a long discussion about its
quantum numbers, the LHCb experiment was recently able to determine parity and total 
angular momentum, assigning the quantum numbers $J^{PC}=1^{++}$ to the $X(3872)$ \cite{Aaij:2013zoa}.

The quantum numbers and the proximity of the mass of the $X(3872)$ to the $\bar{D}^0D^{*0}$ threshold suggest its interpretation
as a loosely-bound $S$-wave hadronic molecule of $D^{(*)}$ mesons with the particle content
\cite{Close:2003sg,Pakvasa:2003ea,Voloshin:2003nt,Wong:2003xk,Braaten:2003he,Swanson:2003tb}
\begin{equation}
	X=\frac{1}{\sqrt{2}}\l(\bar{D}^0D^{*0}+D^0\bar{D}^*{}^0\r).
	\label{eq:Xpartcontent}
\end{equation}
The binding energy of the molecule, $E_X$, is then given as the difference of the sum of the masses of the $D^0$ and $D^{*0}$ meson, 
$m_D$ and $m_{D^*}$, and the mass of the $X(3872)$, $M_X$. Using the latest values from the review of
particle properties \cite{Beringer:1900zz}, we obtain
\begin{equation}
	E_X=m_{D^*}+m_{D}-M_X=(0.17\pm 0.26)\tn{ MeV}.
	\label{eq:mxmdmdst}
\end{equation}
This energy $E_X$ is small compared to the natural energy scale set by one-pion
exchange, $m_\pi^2/(2M_{DD^*})\approx 10$ MeV, where $M_{DD^*}$ is the reduced mass of the $D^0$ and $D^{*0}$ mesons. Thus
the $X(3872)$ displays universal properties determined by its small binding energy or, equivalently, the large 
$\bar{D}^0 D^{*0}$ $S$-wave scattering length $a_s= \sqrt{2M_{DD^*}E_X}$ \cite{Braaten:2004rn}. 
The exploration of this universality for the $X$ using effective field theory methods was initiated
by Braaten and Kusunoki \cite{Braaten:2003he}. A number of predictions for production amplitudes,
decays, formation, and line shapes of the $X(3872)$ were obtained within this framework (see Ref.~\cite{Braaten:2009zz} for
a review). The influence of three-body $D\bar{D}\pi$ interactions on the properties of the $X(3872)$ was found to be 
moderate in a Faddeev approach \cite{Baru:2011rs}.
Finally, we note that universality also determines the interactions of the $X(3872)$ with neutral $D$ and $D^{*}$ mesons
\cite{Canham:2009zq}. 

The corrections to universality can be calculated systematically using an effective field theory for the $X$
with explicit pions, called XEFT, which was developed by Fleming, Kusunoki, Mehen and van Kolck in 2007 
\cite{Fleming:2007rp}. They applied XEFT to calculate the partial decay width $\Gamma\l[X\rightarrow D^0\bar{D}^0\pi^0\r]$ at next-to-leading 
order (NLO) in the XEFT power counting. Later, their work was extended to describe hadronic decays of the $X(3872)$
to the $\chi_{cJ}$ \cite{Fleming:2011xa}. In Ref.~\cite{Braaten:2010mg}, it was pointed out that
XEFT can also be extended to systems with an additional pion with energies close to the $D^* \bar{D}^*$ threshold.
As an example, the cross section for the breakup of the $X$ into $D^{*+}\bar{D}^{*0}$ in the scattering of a 
low-energy charged pion was calculated.

In its structure, XEFT is similar to the Kaplan-Savage-Wise (KSW) theory for nucleon-nucleon ($NN$) interactions \cite{Kaplan:1998tg}. Short-range interactions are parametrized via contact terms and medium- and long-range interactions are given by 
pion exchanges, which are treated perturbatively.
A striking feature of XEFT is that the expansion parameter for the pions is small compared to that in KSW theory for nucleons where large corrections occur at next-to-next-to-leading order in KSW power counting and the perturbative treatment of pions fails \cite{Fleming:1999ee}. Furthermore, since the hyperfine splitting of the $D^0$ and $D^{*0}$ is only about 7 MeV larger than the neutral pion mass, pions as well as the $D^0$ and $D^{*0}$ mesons
are treated non-relativistically. States containing
charged $D^{(*)}$ mesons, such as $D^*{}^+D^-$, are integrated out, since they lie about $8\tn{ MeV}$ above the threshold 
for neutral $D^{(*)}$ mesons. If they are included in the theory, they 
occur first at next-to-next-to-leading order (NNLO) in the power counting \cite{Fleming:1999ee}.

Ultimately, it should be possible to understand the peculiar nature of the $X(3872)$ directly from QCD.
In Ref.~\cite{Yang:2012mya}, some evidence against the quantum number assignment $J^{PC}=2^{-+}$ from a quenched 
lattice calculation was provided.  The first lattice results for the $X$ in full QCD were recently published 
by Prelovsek and Leskovec in \cite{Prelovsek:2013cra}. They found a candidate for the $X(3872)$ about $11\pm7\tn{ MeV}$ 
below the $\bar{D}^0D^{*0}$ threshold. Their simulation was performed on a relatively small lattice with a box length
of approximately $2\tn{ fm}$ and up and down quark masses of about four times the physical value. 
These first results underscore the importance to understand the dependence of the properties
of the $X$ on the volume and the light quark masses. 

Wang and Wang used a unitarized heavy-hadron chiral perturbation theory
approach with pion exchange and a contact interaction in the channel of the $X$ \cite{Wang:2013kva}.
They claimed that
the properties of the $X(3872)$ are insensitive to the contact interaction and concluded
that the binding energy of the $X$ and its quark mass dependence 
are determined by pion exchange alone. This conclusion was challenged by
Baru et al., who investigated the quark mass dependence within the framework of a non-relativistic 
Faddeev-type three-body equation with non-perturbative pions \cite{Baru:2013rta}.  They included a 
$D\bar{D}\pi$ contact interaction to render the equation well defined
and found that the binding energy of the $X$ is indeed sensitive to the quark mass dependence of this term.

In this work, we calculate the light quark mass dependence of the $X(3872)$ in XEFT where pions are perturbative.
To the order we are working, the quark mass dependence is synonymous to
the pion mass dependence because of the Gell-Mann-Oakes-Renner relation \cite{GellMann:1968rz}:
\begin{equation}
m_\pi^2 = -(m_u + m_d) \langle 0 | \bar{u} u + \bar{d}d| 0 \rangle /f^2\,,
\label{eq:mq-mpi}
\end{equation}
where $f\approx 130$ MeV is the pion decay constant and 
$\langle 0 | \bar{u} u | 0 \rangle = \langle 0 | \bar{d} d | 0 \rangle = (-283(2) \mbox{ MeV})^3$ 
is the light quark condensate in the $\overline{MS}$ scheme at 2 GeV \cite{McNeile:2012xh}. 
In the following, we will therefore refer
only to the pion mass dependence which is more convenient for 
our purpose and treat the pion mass as a parameter that can be
varied by adjusting the values of the quark masses.

The paper is organized as follows: In Sec.~\ref{sec:ddstartrans}, we review XEFT and discuss the 
diagrams contributing to $\bar{D}^0D^{*0}$ scattering in the $C=+1$ channel up to NLO.\footnote{For 
simplicity, we will from now on refer to this channel as the $X$ channel.} The issue of 
infrared divergences arising from on-shell pions and their treatment is discussed in Sec.~\ref{sec:irdiv}. Our results
for the binding energy of the $X$ and scattering length in the $X$ channel are given
in Sec.~\ref{sec:bindscattleng}. In Sec.~\ref{sec:chiralres}, we discuss the chiral extrapolations of
the binding energy and scattering length in detail. Finally, we present our conclusions and an outlook
on future work in Sec.~\ref{sec:conc}.

\section{XEFT and the $\bar{D}^0D^{*0}$ scattering amplitude}
\label{sec:ddstartrans}
The Lagrangian for XEFT was derived in Ref.~\cite{Fleming:2007rp} from heavy-hadron chiral perturbation theory. 
It contains non-relativistic fields for the $D^0$, $D^{*0}$, $\bar{D}^0$, and $\bar{D}^{*0}$ mesons
as well as non-relativistic pion fields.  The charged $D$ and $D^*$ mesons have been integrated out of the theory.
This can be done if one is interested in physics near the neutral threshold because the charged threshold
is about $8\tn{ MeV}$ higher in energy. This implies that typical momenta of charged mesons are of the order $120\tn{ MeV}$, 
being much larger than the typical momenta of the neutral mesons which are in the order of the binding momentum $\sim 20\tn{ MeV}$ 
for $E_X=0.2\tn{ MeV}$. If the charged states are not integrated out, they appear at NNLO in the power counting. 
Here, we work only to NLO.
The interaction between the  $D$ and $D^*$ mesons is given by pion exchange and by contact interactions in the 
$C=+1$ channel for $\bar{D}^0D^{*0}$ scattering. The Lagrangian reads
\begin{align}
  \mathcal{L}=&\boldsymbol{D}^\dagger\l(i\partial_0+\frac{\overrightarrow{\nabla}^2}{2m_{D^*}}\r)
  \boldsymbol{D}+D^\dagger\l(i\partial_0+\frac{\overrightarrow{\nabla}^2}{2m_D}\r)D\notag\\
  +&\boldsymbol{\bar{D}}^\dagger\l(i\partial_0+\frac{\overrightarrow{\nabla}^2}{2m_{D^*}}\r)
  \boldsymbol{\bar{D}}+\bar{D}^\dagger\l(i\partial_0+\frac{\overrightarrow{\nabla}^2}{2m_D}\r)\bar{D}+\pi^\dagger\l(
  i\partial_0+\frac{\overrightarrow{\nabla}^2}{2m_\pi}+\delta\r)\pi\notag\\
  +&\frac{g}{\sqrt{2}f}\frac{1}{\sqrt{2m_\pi}}\l(D\boldsymbol{D}^\dagger\cdot\overrightarrow{\nabla}\pi+\bar{D}^\dagger
  \boldsymbol{\bar{D}}\cdot\overrightarrow{\nabla}\pi^\dagger\r)+\tn{h.c.}\notag\\
  -&\frac{C_0}{2}\l(\boldsymbol{\bar{D}}D+\boldsymbol{D}\bar{D}\r)^\dagger\cdot\l(\boldsymbol{\bar{D}}D+
  \boldsymbol{D}\bar{D}\r)\notag\\
  +&\frac{C_2}{16}\l(\boldsymbol{\bar{D}}D+\boldsymbol{D}\bar{D}\r)^\dagger\cdot\l(
  \boldsymbol{\bar{D}}\overleftrightarrow{\nabla}^2D+\boldsymbol{D}\overleftrightarrow{\nabla}^2\bar{D}\r)+\tn{h.c.}\notag\\
  -&\frac{D_2\mu^2}{2}\l(\boldsymbol{\bar{D}}D+\boldsymbol{D}\bar{D}\r)^\dagger\cdot\l(\boldsymbol{\bar{D}}D+
  \boldsymbol{D}\bar{D}\r)+\ldots\, ,
  \label{eq:xeftlagrangian}
\end{align}
where $\overleftrightarrow{\nabla}\equiv\overleftarrow{\nabla}-\overrightarrow{\nabla}$ and the ellipsis denote 
higher order interactions.  The Lagrangian \eqref{eq:xeftlagrangian} is invariant under charge conjugation, parity inversion,
time reversal, and exhibits Galilean invariance.  Moreover,  $m_\pi=135\tn{ MeV}$ is
the neutral pion mass, $g=0.5$ is the $D$ meson axial coupling constant,  $f=132\tn{ MeV}$ is
the pion decay constant, and $\Delta=m_{D^*}-m_D=142\tn{ MeV}$ is 
the hyperfine splitting of the $D^0$ and $D^{*0}$ mesons. The mass scales $\mu$ and $\delta$ are defined as $\mu^2\equiv\Delta^2-m_\pi^2=\l(44\tn{ MeV}\r)^2$
and $\delta\equiv\Delta-m_\pi=7\tn{ MeV}$. We treat $\delta$ as a small mass scale compared to the pion mass and expand amplitudes 
in $\delta/m_\pi$ (cf. Appendix \ref{sec:sigmaaoVI}). The coupling constants $C_0$, $C_2$, and $D_2$ and the power counting in XEFT are 
discussed below. To NLO, the light quark mass dependence enters via the pion exchange and the contact interaction $D_2$. 

An essential feature of XEFT is the perturbative approach to include pions. By means of naive dimensional analysis, 
the two-pion exchange is suppressed compared to the one-pion exchange by a factor of
\begin{equation}
	\frac{g^2M_{DD^*}\mu}{4\pi f^2}\approx\frac{1}{20},
	\label{eq:expfact}
\end{equation}
which justifies the perturbative treatment of pions \cite{Fleming:2007rp}.
Besides, we expand all graphs in powers of the mass ratios
$m_\pi/m_D\approx0.07$ and $\delta/m_\pi\approx0.04$, which are of the same order as the expansion parameter in \eqref{eq:expfact}. Hence, diagrams including a pion with an additional suppression factor of $m_\pi/m_D$ or $\delta/m_\pi$ are in the same order of magnitude as the two-pion exchange graph which is of NNLO and can be neglected.\footnote{
Note that for the calculation of the decay width of the $X(3872)$, neglecting $m_\pi/m_D$ terms is not a good approximation. 
We will come to this issue in chapter \ref{sec:impart}.}

We have explicitly checked that corrections from relativistic pions can be neglected at NLO.
Expanding these contributions around the non-relativistic limit, we find that relativistic corrections to NLO diagrams 
including pions are suppressed by an additional power of $m_\pi/m_D$.

Since we treat the $X(3872)$ as an $S$-wave hadronic molecule, the total angular momentum is given by the $D^{*0}$ meson's spin. 
From angular momentum conservation thus follows that the polarizations of the incoming and outgoing $D^{*0}$ mesons have to coincide. 
Using spin indices $i$ and $j$, the spin dependence of the transition amplitude is of the form
\begin{equation}
	\hat{\mathcal{A}}_{ij}=\delta_{ij}\mathcal{A}\,.
	\label{eq:spindepamp}
\end{equation}
For the discussion of the binding energy and scattering length, it is sufficient to consider the scalar amplitude $\mathcal{A}$.

In XEFT power counting, the binding momentum, the $D^0$ and $D^{*0}$ meson's as well as the pion's typical momentum and the scale $\mu$ are counted as order $Q$. All propagators are of order $Q^{-2}$ and loops are of order $Q^5$ \cite{Fleming:2007rp}.
At leading-order (LO) $Q^{-1}$, there is only one contact term, $C_0$. Note also that 
appending a loop ($Q^{5}$) including two $D^{(*)}$ meson propagators ($Q^{-4}$) and a $C_0$ vertex ($Q^{-1}$) 
does not change the order of any diagram. Therefore, $C_0$ vertices have to be resummed to all orders.
\begin{figure}[ht]
	\begin{center}
		\includegraphics[width=0.7\textwidth]{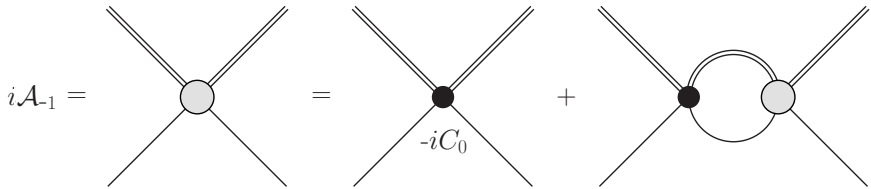}
	\end{center}
	\caption{Leading order contributions to the $\bar{D}^0 D^{*0}$ scattering amplitude. The $\bar{D}^0$ and the $D^{*0}$ mesons are indicated by single and double lines, respectively.} 
	\label{fig:LOcont}
\end{figure}
Using the power divergence subtraction procedure (PDS) \cite{Kaplan:1998tg}, the LO
amplitude at energy $E$ for $\bar{D}^0D^{*0}$ scattering in the $C=+1$ channel, 
depicted in Fig.~\ref{fig:LOcont}, is given as 
\begin{equation}
	i\mathcal{A}_{-1}=\frac{2\pi i}{M_{DD^*}}\frac{1}{-\gamma+\sqrt{-2M_{DD^*}E-i\epsilon}},
	\label{eq:am1}
\end{equation}
where $M_{DD^*}$ is the reduced mass of the $D^0$ and $D^{*0}$ mesons. The quantity $\gamma$ is defined as
\begin{equation}
	\gamma\equiv\frac{2\pi}{M_{DD^*}C_0(\Lambda)}+\Lambda,
	\label{eq:defgamma}
\end{equation}
with the PDS renormalization scale $\Lambda$. Taking $\Lambda$ of order $Q$, we see that the LO amplitude indeed scales as $Q^{-1}$. It has a pole at the LO binding energy $E_X^\tn{LO}=\gamma^2/(2M_{DD^*})$.\footnote{After the inclusion of pions, the pole position of the scattering amplitude becomes complex and will be denoted by $B$. The binding energy, $E_X$, is then given as the real part of $B$.} Hence, $\gamma$ can be identified with the LO binding momentum.

\begin{figure}[ht]
	\begin{center}
		\includegraphics[width=0.8\textwidth]{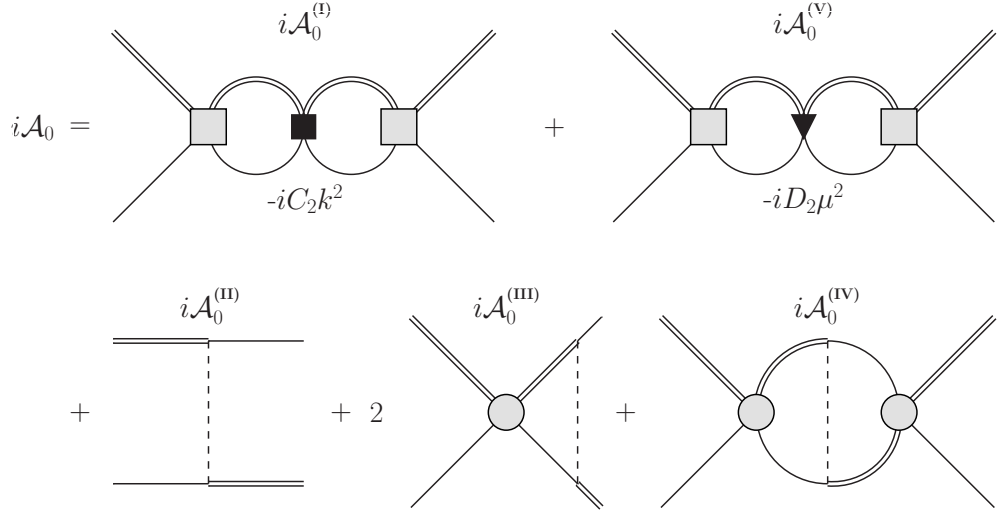}\\
	\end{center}
	\begin{flushleft}
		where
	\end{flushleft}
	\begin{center}
		\includegraphics[width=0.57\textwidth]{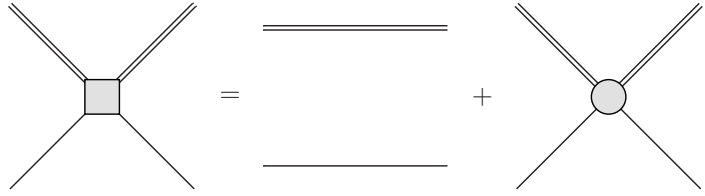}
		\caption{Next-to-leading order contributions to the $\bar{D}^0 D^{*0}$ scattering amplitude. The pions, the $\bar{D}^0$, and the $D^{*0}$ mesons are indicated by dashed lines, solid lines, and double lines, respectively.}\label{fig:NLOcont}
	\end{center}
\end{figure}

At order $Q^0$, which is NLO in XEFT power counting, we have to include further contact interactions with coupling constants $C_2$ and $D_2$ which are both of order $Q^{-2}$. In the XEFT Lagrangian \eqref{eq:xeftlagrangian} the coupling constant $C_2$ comes with two derivatives and $D_2$ with a factor of $\mu^2$. Note that this is different from the factor $m_\pi^2$ in KSW counting because in XEFT
the typical momenta of the $D^{(*)}$ mesons are of order $\mu \ll m_\pi$.
Vertices including $C_2$ or $D_2$ thus are of order $Q^0$. For each pion exchange there are two $D^0D^{*0}\pi^0$ axial couplings, each of order $Q^1$, and one pion propagator of order $Q^{-2}$ resulting in order $Q^0$, too. We end up with the five diagrams shown in Fig.~\ref{fig:NLOcont}.

\begin{figure}[ht]
	\begin{center}
		\includegraphics[width=0.4\textwidth]{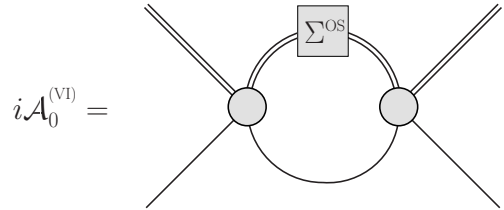}
	\end{center}
	\caption{NLO contribution from $D^{*0}$ self energy. We use the same notation like in Fig.~\ref{fig:NLOcont}.}
	\label{fig:dstbub}
\end{figure}

\begin{figure}[ht]
	\begin{center}
		\includegraphics[width=0.8\textwidth]{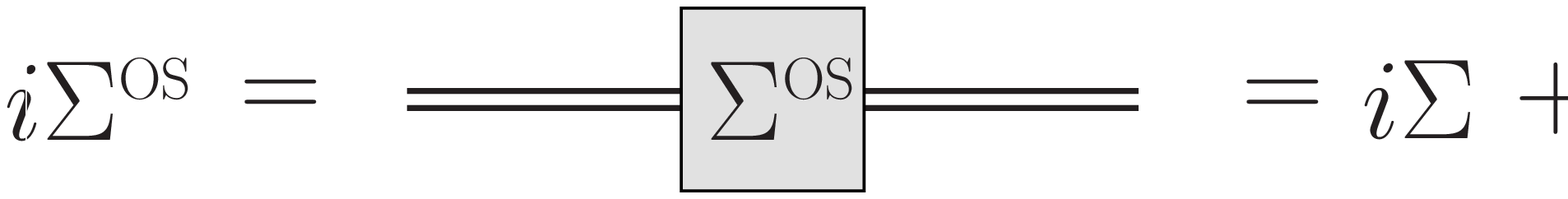}
	\end{center}
	\caption{Self energy graph and counter term for the $D^{*0}$. We use the same notation like in Fig.~\ref{fig:NLOcont}.}
	\label{fig:dstse}
\end{figure}

A novel feature of XEFT is the occurrence of a sixth diagram, $\mathcal{A}_0^\tn{(VI)}$, depicted in Fig.~\ref{fig:dstbub}. 
The transition amplitude $\mathcal{A}_0^\tn{(VI)}$ comes from the self energy diagram for the $D^{*0}$ shown in 
Fig.~\ref{fig:dstse}. It does not occur in KSW theory due to the different kinematics for nucleons. Since pions are always off-shell for the $NN$ system, the bare self energy diagram is purely imaginary and removed by the counter term in the on-shell renormalization scheme. We will further discuss this issue in section \ref{sec:irdiv}. 
We denote the external incoming momenta in the center-of-momentum frame of the $D^0$ and $D^{*0}$ mesons $\mathbf{p}$ and $-\mathbf{p}$ and the outgoing $\mathbf{l}$ and $-\mathbf{l}$, respectively, use $p=\l|\mathbf{p}\r|=\l|\mathbf{l}\r|$ for elastic scattering and evaluate the amplitudes at on-shell energy $E=p^2/(2M_{DD^*})$. We start with the contributions of the NLO contact interactions, $\mathcal{A}_0^\tn{(I)}$ and $\mathcal{A}_0^\tn{(V)}$, respectively. Similar to the $NN$ case \cite{Kaplan:1998tg}, 
we acquire 
\begin{subequations}
\begin{align}
	&i\mathcal{A}_0^\tn{(I)}=\frac{-iC_2p^2}{C_0^2}\mathcal{A}^2_{-1},\label{eq:NLOcontC2}\\
	&i\mathcal{A}_0^\tn{(V)}=\frac{-iD_2\mu^2}{C_0^2}\mathcal{A}^2_{-1}.
	\label{eq:NLOcontD2}
\end{align} 
\end{subequations}
The one-pion exchange transition amplitude $\hat{\mathcal{A}}_0^\tn{(II)}{}_{ij}$ is given as
\begin{equation}
	i\hat{\mathcal{A}}_0^\tn{(II)}{}_{ij}=\frac{ig^2}{2f^2}\frac{\l(\boldsymbol{\varepsilon}_i\cdot\mathbf{q}\r)\l(\boldsymbol{\varepsilon}_j\cdot\mathbf{q}\r)}{\mathbf{q}^2-\mu^2},
	\label{eq:a0II}
\end{equation}
where $\mathbf{q}$ is the momentum transfer and $\boldsymbol{\varepsilon}_i$ and $\boldsymbol{\varepsilon}_i$ are the polarization vectors of the $D^{*0}$ mesons. Projecting $\hat{\mathcal{A}}_0^\tn{(II)}{}_{ij}$ on the $X$ channel and factoring out the spin dependence as in \eqref{eq:spindepamp}, we end up with
\begin{equation}
	i\mathcal{A}_0^\tn{(II)}=\frac{ig^2}{6f^2}\l[1+\frac{\mu^2}{4p^2}\log\l(1-\frac{4p^2}{\mu^2}\r)\r].
	\label{eq:a0IIpartwave}
\end{equation}
For the one- and two-loop diagrams with one-pion exchange, $\mathcal{A}_0^\tn{(III)}$ and $\mathcal{A}_0^\tn{(IV)}$, we acquire
\begin{subequations}
\begin{align}
	&i\mathcal{A}_0^\tn{(III)}=\frac{ig^2}{3f^2}\l[\l(ip+\Lambda\r)+i\mu^2\frac{1}{2p}\log\l(1+\frac{2p}{\mu}\r)\r]\frac{M_{DD^*}}{2\pi}\mathcal{A}_{-1},\label{eq:a0III}\\
	&i\mathcal{A}_0^\tn{(IV)}=\frac{ig^2}{6f^2}\l[\l(ip+\Lambda\r)^2+\mu^2\l(\log\l(\frac{\Lambda}{-2ip-i\mu}\r)+1+R\r)\r]\l(\frac{M_{DD^*}}{2\pi}\r)^2\mathcal{A}^2_{-1},
	\label{eq:a0IV}
\end{align}
\end{subequations}
with $R\equiv\frac{1}{2}\l(-\gamma_E+\log\l(\tfrac{\pi}{4}\r)+\tfrac{2}{3}\r)$. Since $\mathcal{A}_0^\tn{(IV)}$ depends logarithmically on $\Lambda$ it is required to include the $\mu^2$-dependent vertex proportional to $D_2$ to ensure that physical results are renormalization scale independent \cite{Kaplan:1998tg}.

\section{Infrared divergences and Full $D^{*0}$ Propagator}
\label{sec:irdiv}
For the calculation of diagram $\mathcal{A}_0^\tn{(VI)}$ in Fig.~\ref{fig:dstbub}, we first consider the renormalized $D^{*0}$ self energy shown in Fig.~\ref{fig:dstse}. Explicit calculations for the bare self energy diagram and $\mathcal{A}_0^\tn{(VI)}$ can be found in Appendix \ref{sec:sigmaaoVI}. We use the on-shell renormalization scheme, where the counter term, $i\delta_\Sigma$, on the right 
hand side of Fig.~\ref{fig:dstse} is chosen such that the real part of the pole position of the $D^{*0}$ propagator stays at the on-shell $D^{*0}$ energy $p_0=p^2/2m_{D^*}$. This implies that the counter term has to cancel the imaginary part of the bare self energy, $i\Sigma$, which is in the PDS renormalization scheme given as\footnote{Note that taking the $D^{*0}$ energy to be of order $p^2/m_{D^*}\sim\l|\Sigma^\tn{os}\r|$ and neglecting the pion loop dressed $D^{*0}$ propagator with additional factors of $m_\pi/m_D$, the bare $D^{*0}$ self energy is energy independent and thus the field strength renormalization constant is $1$. All corrections are suppressed by $g^2M_{DD^*}\mu/\l(4\pi f^2\r)$ times powers of $m_\pi/m_D$.}
\begin{equation}
	i\Sigma=\frac{ig^2}{24\pi f^2}\l(i\mu^3+\Lambda\mu^2\r).
	\label{eq:sigmadst}
\end{equation}
For pion masses $m_\pi>\Delta$, $\mu$ becomes imaginary, such that the bare self energy is imaginary, too. It follows for the counter term
\begin{equation}
	i\delta_\Sigma=\begin{cases}
		-\frac{ig^2}{24\pi f^2}\Lambda\mu^2 &, m_\pi<\Delta,\\
		-\frac{ig^2}{24\pi f^2}\l(i\mu^3+\Lambda\mu^2\r) &, m_\pi>\Delta.
	\end{cases}
	\label{eq:deltam}
\end{equation}
We see that as soon as pions can not go on-shell, the on-shell renormalized self energy yields zero and hence $\mathcal{A}_0^\tn{(VI)}$ vanishes, too. Note that pions in $NN$ scattering are always off-shell, implying that the diagram in Fig.~\ref{fig:dstbub} does not 
contribute in KSW theory. 
However, for the case $m_\pi<\Delta$, we obtain
\begin{equation}
	i\mathcal{A}_0^\tn{(VI)}=\frac{i}{p}2\pi i\Sigma^\tn{OS}\l(\frac{M_{DD^*}}{2\pi}\r)^2\mathcal{A}_{-1}^2,
	\label{eq:a0VI}
\end{equation}
which is infrared divergent.
\begin{figure}[ht]
	\begin{center}
		\includegraphics[width=0.7\textwidth]{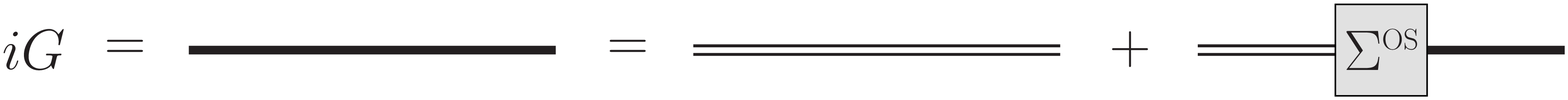}
	\end{center}
	\caption{Full $D^{*0}$ propagator. The free $D^{*0}$ propagator is denoted by a double line, the full $D^{*0}$ propagator by a thick, single line.}
	\label{fig:resdst}
\end{figure}
The divergence occurs due to an inappropriate expansion at low energies. To trace the origin of the infrared divergence, let us consider the full $D^{*0}$ propagator with resummed self energy shown in Fig.~\ref{fig:resdst}
\begin{equation}
	iG=\frac{i}{p_0-p^2/2m_{D^*}+i\epsilon}\l(1+i\Sigma^\tn{OS}iG\r)=\frac{i}{p_0-p^2/2m_{D^*}+\Sigma^\tn{OS}+i\epsilon}.
	\label{eq:resdst}
\end{equation}
For pion masses $m_\pi<\Delta$, $\Sigma^\tn{os}$ is purely imaginary and is related to the decay width of the $D^{*0}$, $\Gamma_*$, by $\Sigma^\tn{os}=i\Gamma_*/2$. Hence, the full propagator takes the nonzero decay width of the $D^{*0}$ into account.

Now, we use \eqref{eq:resdst} to evaluate $\mathcal{A}_0^\tn{(VI)}$ with full instead of free $D^{*0}$ propagators. To avoid double counting, we replace one of the two free $D^{*0}$ propagators in the loop in Fig.~\ref{fig:dstbub} by the full one.
This yields the resummed amplitude $\l(\mathcal{A}_0^\tn{(VI)}\r)^\tn{res}$
\begin{align}
	i\l(\mathcal{A}_0^\tn{(VI)}\r)^\tn{res}=&\frac{-2ip+2\sqrt{-p^2-i\kappa^2-i\epsilon}}{i\kappa^2}2\pi i\Sigma^\tn{OS}\l(\frac{M_{DD^*}}{2\pi}\r)^2\mathcal{A}_{-1}^2\notag\\
	=&\frac{-2+2\sqrt{1+i\kappa^2/p^2}}{i\kappa^2/p^2}\cdot\frac{i}{p}2\pi i\Sigma^\tn{OS}\l(\frac{M_{DD^*}}{2\pi}\r)^2\mathcal{A}_{-1}^2,
	\label{eq:resa0VI}
\end{align}
with $i\kappa^2=2M_{DD^*}\Sigma^\tn{OS}$. As can be seen from the first line, the resummed transition amplitude is infrared finite for all values of $m_\pi$. Let us expand the first factor in the second line of \eqref{eq:resa0VI} around $\kappa^2/p^2=0$, which is equivalent to expanding the full $D^{*0}$ propagator. At zeroth order we reproduce Eq. \eqref{eq:a0VI}. It is clear that this expansion is invalid for momenta $p\lesssim\l|\kappa\r|\approx5\tn{ MeV}$ and we have to use the full $D^{*0}$ propagator instead of the expanded one. Physically speaking, at energies close to the $\bar{D}^0D^{*0}$ threshold, the main contribution to the loop integral comes from the low-energy regime, where the virtual $D^{*0}$ meson can propagate much longer than the $D^{*0}$'s average lifetime. Therefore, it is not justified to treat its decay in perturbation theory anymore.

When dressing the $D^{*0}$ propagators in diagrams $\mathcal{A}_0^\tn{(I)}$ to $\mathcal{A}_0^\tn{(V)}$ with pion loops, similar infrared divergences occur. Hence, for consistency, we have to use the full $D^{*0}$ propagator for all these diagrams, as well as for the LO amplitude $\mathcal{A}_{-1}$. Note that the size of $\kappa$ is in the order of the typical momentum scale of XEFT, $Q$. This implies that the full $D^{*0}$ propagator is still of order $Q^{-2}$, i.e. the power counting remains unaltered.

\section{Binding energy and scattering length}
\label{sec:bindscattleng}
The central point of this work is the calculation of low-energy observables for the $X$ in dependence on the light quark masses. In this section we present the results for the transition amplitudes, the binding energy and the scattering length at NLO. We renormalize the transition amplitudes by use of the coupling constants $C_0$, $C_2$, and $D_2$. 
\subsection{Transition amplitudes up to NLO}
As seen in the previous section, amplitudes containing a $D^{*0}$ propagator dressed with a pion loop exhibit infrared divergences. Therefore, considering the low-energy regime, we have to reexpress the LO and NLO amplitudes $\mathcal{A}_{-1}$ and $\mathcal{A}_0^\tn{(I)}$ to $\mathcal{A}_0^\tn{(V)}$ with full $D^{*0}$ propagators. Note that the LO amplitude $\mathcal{A}_{-1}$ with the $D^{*0}$ propagator dressed to all orders automatically includes the amplitude $\mathcal{A}_0^\tn{(VI)}$, which thus must not be taken into account separately to avoid double counting. Skipping the superscript for convenience, the amplitudes read
\begin{subequations}
\begin{align}
	&i\mathcal{A}_{-1}=\frac{2\pi i}{M_{DD^*}}\frac{1}{-\gamma+\eta},\label{eq:am1res}\\
	&i\mathcal{A}_0^\tn{(I)}=\frac{-iC_2}{C_0^2}\l(p^2+2M_{DD^*}\Sigma^\tn{os}\frac{-\eta+\Lambda}{-\gamma+\Lambda}\r)\mathcal{A}_{-1}^2,\label{eq:a0Ires}\\
	&i\mathcal{A}_0^\tn{(II)}=\frac{ig^2}{6f^2}\l(1+\frac{\mu^2}{4p^2}\log\l(1-\frac{4p^2}{\mu^2}\r)\r),\label{eq:a0IIres}\\
	&i\mathcal{A}_0^\tn{(III)}=\frac{ig^2}{3f^2}\l(\l(-\eta+\Lambda\r)+\frac{i\mu^2}{2p}\log\l(1+\frac{2p}{i\eta+\mu-p}\r)\r)\frac{M_{DD^*}}{2\pi}\mathcal{A}_{-1},\label{eq:a0IIIres}\\
	&i\mathcal{A}_0^\tn{(IV)}=\frac{ig^2}{6f^2}\l(\l(-\eta+\Lambda\r)^2+\mu^2\l(\log\l(\frac{\Lambda}{2\eta-i\mu}\r)+1+R\r)\r)\l(\frac{M_{DD^*}}{2\pi}\r)^2\mathcal{A}_{-1}^2,\label{eq:a0IVres}\\
	&i\mathcal{A}_0^\tn{(V)}=\frac{-iD_2\mu^2}{C_0^2}\mathcal{A}_{-1}^2,\label{eq:a0Vres}
\end{align}
\end{subequations}
with $\eta$ defined as $\eta\equiv\sqrt{-p^2-2M_{DD^*}\Sigma^\tn{os}-i\epsilon}$. All diagrams are finite for $p\rightarrow0$ for all values of $m_\pi$. For $\Sigma^\tn{os}\rightarrow 0$, $\eta\rightarrow-ip$ and the results for the diagrams with the free $D^{*0}$ propagator are reproduced. The LO diagram $\mathcal{A}_{-1}$ has a pole at $p^2=\eta_B^2\equiv-\gamma^2-2M_{DD^*}\Sigma^\tn{os}$, corresponding to an LO pole position at $-E=B^\tn{LO}=\gamma^2/(2M_{DD^*})+\Sigma^\tn{os}$, which is complex for $m_\pi<\Delta$. The binding energy is given as the real part of the pole position and can be adjusted with $\gamma$ for renormalization.\footnote{Since we are not considering inelastic channels like for example $X\rightarrow J/\Psi\pi^+\pi^-$, the LO binding momentum, $\gamma$, is real valued \cite{Stapleton:2009ey}.} Our result for the LO transition amplitude $\mathcal{A}_{-1}$ is in agreement with the results from \cite{Stapleton:2009ey}, where the authors obtained the $\bar{D}^0D^{*0}$ transition amplitude to LO by analytically continuing the parameters of a threshold resonance form for two stable particles to the complex plane.
\subsection{Renormalization of the transition amplitude}
\label{sec:renor}
The transition amplitude $\mathcal{A}_0=\mathcal{A}_0^\tn{(I)}+\mathcal{A}_0^\tn{(II)}+\mathcal{A}_0^\tn{(III)}+\mathcal{A}_0^\tn{(IV)}+\mathcal{A}_0^\tn{(V)}$ has to be renormalization scale independent up to NLO. It follows for the coupling constants $C_2$ and $D_2$
\begin{subequations}
\begin{align}
	&C_2=\frac{M_{DD^*}}{2\pi}\frac{r_0}{2}\l(C_0\r)^2\equiv c_2\l(C_0\r)^2,\label{eq:renorc2}\\
	&D_2=\frac{6f^2}{g^2}\l(\frac{2\pi}{M_{DD^*}}\r)^2\l(d_2+\log\l(\frac{\Lambda}{\mu^\tn{ph}}\r)-R\r) \l(C_0\r)^2\label{eq:renord2},
\end{align}
\end{subequations}
in analogy to \cite{Kaplan:1998tg} and \cite{Fleming:2007rp}. Here and in the following, the superscript ph denotes the physical value of a quantity, i.e. at the physical pion mass. The quantity $r_0$ with dimension of length can be identified with the effective range in the pionless theory. We further absorbed the constant $R$, which occurs in PDS, in the coupling constant $D_2$. Following the arguments  of \cite{Fleming:2007rp}, we use $r_0\in \l[0,1/100\tn{MeV}\r]$, such that the maximum value of $r_0$ is inversely proportional to the momentum scales integrated out of the theory. For the dimensionless parameter $d_2$, we use that the numerical value of the terms in the parentheses with $\mu^2$ as prefactor in Eq. \eqref{eq:a0IV} is about $0.9$ evaluated at the physical pion mass with $\Lambda\sim\mu^\tn{ph}$. We will therefore take $d_2\in \l[-1,1\r]$.

\subsection{The binding energy at NLO}
\label{sec:deltab}
To calculate the binding energy at NLO, we compare the XEFT result for the transition amplitude $\mathcal{A}=\mathcal{A}_{-1}+\mathcal{A}_0+\ldots$ with a generic, non-perturbative representation $\mathcal{A}^\tn{np}=Z/\l(E+B\r)+\ldots$ with shifted pole position $B$ and field strength renormalization constant $Z$. The shifts for the pole position and the field strength renormalization constant can then be acquired by expanding both, the XEFT and the non-perturbative expression, around the LO pole position and matching coefficients afterwards. First, we expand the LO amplitude around the LO position of the pole and discard all terms regular at $E=-B^\tn{LO}$, represented by ellipsis 
\begin{equation}
	\mathcal{A}_{-1}=\frac{Z_{-1}}{E+B^\tn{LO}}+\ldots\, .
	\label{eq:am1atpole}
\end{equation}
The LO field strength renormalization constant, $Z_{-1}$, is given as the residue at $E=-B^\tn{LO}$
\begin{equation}
	\Rightarrow\l(Z_{-1}\r)^{-1}=\l[i\frac{\partial}{\partial E}\frac{1}{i\mathcal{A}_{-1}}\r]_{E=-B^\tn{LO}}=\frac{-\l(M_{DD^*}\r)^2}{2\pi}\frac{1}{\gamma}.
	\label{eq:am1resZ}
\end{equation}
Next, we expand the amplitude $\mathcal{A}_0$ around the LO pole position. It divides into three parts: one proportional to $\mathcal{A}_{-1}$, one proportional to $\mathcal{A}_{-1}^2$ and one being finite at $E=-B^\tn{LO}$ again denoted by ellipsis
\begin{align}
	\mathcal{A}_0=&\l[c_2\frac{2\pi}{M_{DD^*}}\frac{\eta_0^2}{\gamma-\Lambda}+\frac{g^2}{6f^2}\frac{M_{DD^*}}{2\pi }\frac{i\mu^2}{\eta_B}\log\l(1+\frac{2\eta_B}{i\gamma+\mu-\eta_B}\r)\r]\mathcal{A}_{-1}+\notag\\
	+&\l[c_2\gamma^2+\frac{g^2}{6f^2}\l(\frac{M_{DD^*}}{2\pi}\r)^2\l[\l(\gamma-\Lambda\r)^2+\mu^2\l(-d_2+\log\l(\frac{\mu^\tn{ph}}{2\gamma-i\mu}\r)+1\r)\r]\r]\mathcal{A}_{-1}^2+\ldots\notag\\
	\equiv& s_1\cdot\frac{Z_{-1}}{E+B^\tn{LO}}+s_2\cdot\frac{Z_{-1}^2}{\l(E+B^\tn{LO}\r)^2}+\ldots\, ,
	\label{eq:splitA0}
\end{align}
where $\eta_0\equiv\sqrt{-2M_{DD^*}\Sigma^\tn{os}}$. The terms linear in $\mathcal{A}_{-1}$ arise due to cancellations in the numerator 
at $E=-B^\tn{LO}$. We now compare the non-perturbative expression $\mathcal{A}^\tn{np}$ at energy $B=B^\tn{LO}+\Delta B$ with field strength renormalization constant $Z=Z_{-1}+\Delta Z$ to the XEFT result $\mathcal{A}$ 
\begin{subequations}
\begin{align}
	\mathcal{A}=\mathcal{A}_{-1}+\mathcal{A}_0+\ldots=&\frac{Z_{-1}+s_1Z_{-1}}{E+B^\tn{LO}}+\frac{s_2Z_{-1}^2}{\l(E+B^\tn{LO}\r)^2}+\ldots\, ,\\
	\mathcal{A}^\tn{np}=\frac{Z}{E+B}+\ldots=&\frac{Z_{-1}+\Delta Z}{E+B^\tn{LO}}-\frac{Z\Delta B}{\l(E+B^\tn{LO}\r)^2}+\ldots\, .
	\label{eq:nppcmp}
\end{align}
\end{subequations}
The shift of the field strength renormalization constant and the pole position to NLO can then be read off by equating the corresponding coefficients
\begin{subequations}
\begin{align}
	\Delta Z^\tn{NLO}=&s_1 Z_{-1},\\
	\Delta B^\tn{NLO}=&-\frac{Z_{-1}^2}{Z_{-1}+\Delta Z}s_2=-\frac{Z_{-1}}{1+s_1}s_2.
	\label{eq:ZBshift}
\end{align}
\end{subequations}
The superscript denotes that these relations hold up to NLO. In summary, the binding energy $E_X$ is given as
\begin{subequations}
	\begin{align}
		&E_X=\operatorname{Re}\l[B^\tn{LO}+\Delta B^\tn{NLO}\r],\label{eq:EXfromB}\\
		&B^\tn{LO}+\Delta B^\tn{NLO}=\frac{\gamma^2}{2M_{DD^*}}+\Sigma^\tn{os}-\frac{Z_{-1}}{1+s_1}s_2,\label{eq:BLOpNLO}\\
		&Z_{-1}=-\frac{2\pi\gamma}{\l(M_{DD^*}\r)^2},\label{eq:Zm1}\\
		&s_1=c_2\frac{2\pi}{M_{DD^*}}\frac{\eta_0^2}{\gamma-\Lambda}+\frac{g^2}{6f^2}\frac{M_{DD^*}}{2\pi }\frac{i\mu^2}{\eta_B}\log\l(1+\frac{2\eta_B}{i\gamma+\mu-\eta_B}\r),\label{eq:s1}\\
		&s_2=c_2\gamma^2+\frac{g^2}{6f^2}\l(\frac{M_{DD^*}}{2\pi}\r)^2\l[\l(\gamma-\Lambda\r)^2+\mu^2\l(-d_2+\log\l(\frac{\mu^\tn{ph}}{2\gamma-i\mu}\r)+1\r)\r].\label{eq:s2}
	\end{align}
	\label{eq:EXNLO}
\end{subequations}
For renormalization, we fix it at the physical value of the pion mass and use $E_X^\tn{ph}=0.2\tn{ MeV}$. This defines a relation between the LO binding momentum $\gamma$, i.e. $C_0$, and the coefficients $c_2$ and $d_2$. 

\subsection{The imaginary part of the pole position}
\label{sec:impart}
\begin{figure}[ht]
	\begin{center}
		\includegraphics[width=0.7\textwidth]{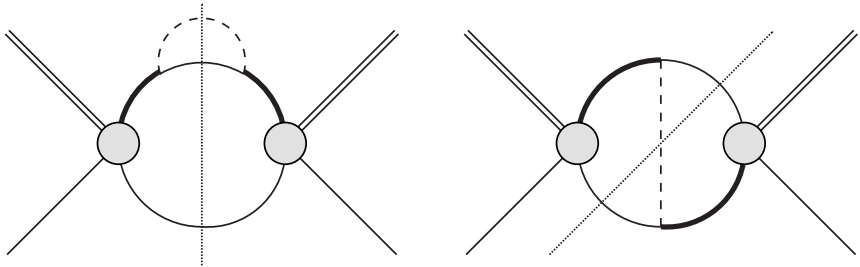}
	\end{center}
	\caption{Cut diagrams, determining the imaginary part of the pole position $B$. We use the same notation as in the previous figures. The cuts through the pion, $D^0$ and $\bar{D}^0$ meson propagators are indicated by dotted lines.}
	\label{fig:cutdiagrams}
\end{figure}

The imaginary part of the pole position $B=B^\tn{LO}+\Delta B^\tn{NLO}$ can be obtained by cutting the diagrams in Figs.~\ref{fig:LOcont} and \ref{fig:NLOcont} with the free $D^{*0}$ propagator replaced with the full one. Consider for example the cut diagrams shown in Fig.~\ref{fig:cutdiagrams}. Note that the first diagram is included in the LO amplitude $\mathcal{A}_{-1}$. Applying the cuts by replacing all cut propagators with appropriate delta distributions and keeping all $m_\pi/m_D$ suppressed terms, the expressions for the imaginary parts coincide with the decay diagrams in \cite{Stapleton:2009ey} at energy $E_X$.
In \cite{Fleming:2007rp}, the authors pointed out that dropping $m_\pi/m_D$ terms is equivalent to treating pions in potential approximation.\footnote{For potential pions, the kinetic energy is much smaller than the pion momentum \cite{Mehen:1999hz}. Dropping $m_\pi/m_D$ terms implies that the kinetic energy of the pions is neglected and thus pions are treated in potential approximation.} But for the decay diagrams, the pions in the final state are on-shell and the potential approximation is invalid. Hence, for the calculation of the imaginary part of the pole position, which is related to the decay width by $\operatorname{Im}\l[B\r]=\Gamma\l[X\rightarrow D^0\bar{D}^0 \pi^0\r]/2$, all $m_\pi/m_D$ suppressed terms have to be kept and the remaining three-body phase space integral has to be evaluated numerically. It is not to be expected that the decay width is well approximated when treating the final state pions in potential approximation. Further discussions can be found in \cite{Voloshin:2003nt,Fleming:2007rp,Stapleton:2009ey}. 
\subsection{The scattering length at NLO}
In the previous chapters we presented results for the transition amplitudes and derived an expression for the binding energy at LO and NLO. We now turn to the calculation of the scattering length. For this purpose we consider the $S$-matrix which is related to the $S$-wave transition amplitude $\mathcal{A}$ by
\begin{equation}
	S-1=e^{2i\delta_s}-1=i\frac{pM_{DD^*}}{\pi}\mathcal{A},
	\label{eq:scattrans}
\end{equation}
with the $S$-wave scattering phase shift $\delta_s$. To take the inelastic channel $DD^*\rightarrow \bar{D}D\pi$ into account, we allow the scattering phase shift to be complex. Equation \eqref{eq:scattrans} can be rewritten and expanded at low energies in $p^2$ as
\begin{equation}
	p\cot{\delta_s}=ip+\frac{2\pi}{M_{DD^*}\mathcal{A}}=-\frac{1}{a_s}+\frac{1}{2}r_sp^2+\ldots\, ,
	\label{eq:effrangeexp}
\end{equation}
which is known as effective range expansion. The quantity $r_s$ is called $S$-wave effective range and $a_s$ is the $S$-wave scattering length. In the pionless theory, the effective range $r_s$ coincides with $r_0$ in Eq. \eqref{eq:renorc2}. However, after the inclusion of pions, the effective range expansion remains valid only up to order $p^0$ for $m_\pi<\Delta$ ($\mu^2>0$). This can be understood by taking a closer look at the Fourier transform of the one-pion exchange in potential approximation
\begin{equation}
	\frac{g^2}{2f^2}\frac{\l(\boldsymbol{\varepsilon}_i\cdot\mathbf{q}\r)\l(\boldsymbol{\varepsilon}_j\cdot\mathbf{q}\r)}{\mathbf{q}^2-\mu^2}\xrightarrow{\tn{F.T.}}\frac{g^2}{8\pi f^2}\l(\boldsymbol{\varepsilon}_i\cdot\boldsymbol{\varepsilon}_j-3\l(\boldsymbol{\varepsilon}_i\cdot\mathbf{\hat{r}}\r)\l(\boldsymbol{\varepsilon}_j\cdot\mathbf{\hat{r}}\r)\r)\frac{\cos\l(\mu r\r)+\mu r\sin\l(\mu r\r)}{r^3}+\ldots\, ,
	\label{eq:ftope}
\end{equation}
occurring in all diagrams involving pions. In contrast to the exponentially decreasing potential for the one-pion exchange as an effective $NN$ interaction, the potential \eqref{eq:ftope} is oscillatory and hence the effective range $r_s$ is not defined \cite{Suzuki:2005ha}. This results in the emergence of terms linear in $p$ when expanding the transition amplitude around $p=0$. Nevertheless, the $S$-wave scattering length is well-defined and can be extracted.

To do this, we take the limit $p\rightarrow0$ of the transition amplitude. At NLO we acquire
\begin{align}
	a_s=&\frac{-M_{DD^*}}{2\pi}\l(\mathcal{A}_{-1}+\mathcal{A}_0\r)\notag\\
	=&\frac{1}{\gamma-\eta_0}-\frac{1}{\l(\gamma-\eta_0\r)^2}\biggl[\frac{r_0}{2}\eta_0^2\frac{\eta_0-\Lambda}{\gamma-\Lambda}+\frac{M_{DD^*}}{2\pi}\frac{g^2}{6f^2}\biggl(\l(\gamma-\Lambda\r)^2-\l(\gamma-\eta_0\r)^2+\notag\\
	+&2\mu^2\frac{\gamma-\eta_0}{i\mu-\eta_0}+\mu^2\l(-d_2+\log\l(\frac{\mu^\tn{ph}}{2\eta_0-i\mu}\r)+1\r)\biggr)\biggr].
	\label{eq:scl}
\end{align}
The scattering length is complex for $m_\pi<\Delta$ and real as soon as the hyperfine splitting of the $D^0$ and $D^{*0}$ mesons is smaller than the neutral pion mass.

\section{Chiral Extrapolations and Results}
\label{sec:chiralres}
For the determination of the quark mass dependence of the binding energy of the $X$, we need the chiral extrapolations of the pion decay constant, the $D$ meson axial coupling constant and the $D^0$ and $D^{*0}$ meson, respectively. We use a superscript $(0)$ to denote the chiral limit value of a quantity. We take the $m_\pi$ dependence of the pion decay constant from \cite{Gasser:1983yg}
\begin{equation}
	f=f^{(0)}\l[1-\frac{1}{4\pi^2{f^{(0)}}^2}m_\pi^2\log\l(\frac{m_\pi}{m_\pi^\tn{ph}}\r)+\frac{\bar{l}_4}{8\pi^2{f^{(0)}}^2}m_\pi^2\r],
	\label{eq:fpchiext}
\end{equation}
with the low-energy constant $\bar{l}_4=4.4$ and $f^{(0)}=124\tn{ MeV}$, implying $f^\tn{ph}=132\tn{ MeV}$ \cite{Gasser:1983yg,Colangelo:2001df}. For the $D$ meson axial coupling constant, we use the recent lattice results from \cite{Becirevic:2012pf}. The chiral extrapolation reads
\begin{equation}
	g=g^{(0)}\l[1-\frac{1+2{g^{(0)}}^2}{4\pi^2{f^{(0)}}^2}m_\pi^2\log\l(\frac{m_\pi}{\mu_\tn{lat}}\r)+\alpha m_\pi^2\r],
	\label{eq:gchiext}
\end{equation}
with the parameters \cite{Becirevic:2012pf}
\begin{equation}
	g^{(0)}=0.46, ~ ~ \alpha=-0.16\tn{ GeV}^{-2}, ~ ~ \mu_\tn{lat}=1\tn{ GeV}.
	\label{eq:latticepara}
\end{equation}
Evaluated at the physical pion mass, the physical value of the $D$ meson axial coupling constant is $g^\tn{ph}=0.5$.

For the quark mass dependence of the $D$ meson masses and hence the hyperfine splitting $\Delta$, we use the results of \cite{Guo:2009ct}
\begin{equation}
	m_{D^{(*)}}=m_{D^{(*)}}^\tn{ph}+\frac{h_1}{m_{D^{(*)}}^\tn{ph}}\l(m_\pi^2-\l(m_\pi^\tn{ph}\r)^2\r),
	\label{eq:mdmdstmqdep}
\end{equation}
with $h_1=0.42$ \cite{Guo:2009ct}. 

	In the KSW theory for $NN$ scattering, the relative size of the two-pion and one-pion exchange graphs is about $1/2$. Due to NNLO coefficients of order $5\sim6$ being much greater than the expansion parameter in KSW theory, contributions at NLO and NNLO are of comparable magnitude and the perturbative treatment of pions fails \cite{Fleming:1999ee}. In XEFT, the two-pion exchange is more strongly suppressed. However, the estimate of the suppression based on \eqref{eq:expfact} 
depends on the quark mass, dominantly through the mass scale $\mu$.
To determine a region of validity for XEFT we use a rather conservative estimate for the upper bound of the expansion parameter and require that the absolute value of \eqref{eq:expfact} is smaller than $0.15$. Even though unnaturally large NNLO coefficients of similar size as in KSW occur, \eqref{eq:expfact} is expected to be small enough to compensate for that and the perturbative inclusion of pions remains valid. For the lower bound we consider that pions are treated non-relativistically in XEFT. We require for the maximum pion velocity $v_\pi\approx\mu/m_\pi\lesssim0.35$. These conditions are fulfilled for $0.98\l(m_\pi^\tn{ph}\r)^2\lesssim m_\pi^2\lesssim2\l(m_\pi^\tn{ph}\r)^2$. Since the coupling constants $c_2$ and $d_2$ are undetermined, it might be that unnaturally large corrections to the LO amplitude occur at NLO. We will come to this issue when discussing the scattering length.

\begin{figure}[ht]
	\begin{center}
		\includegraphics[width=0.7\textwidth]{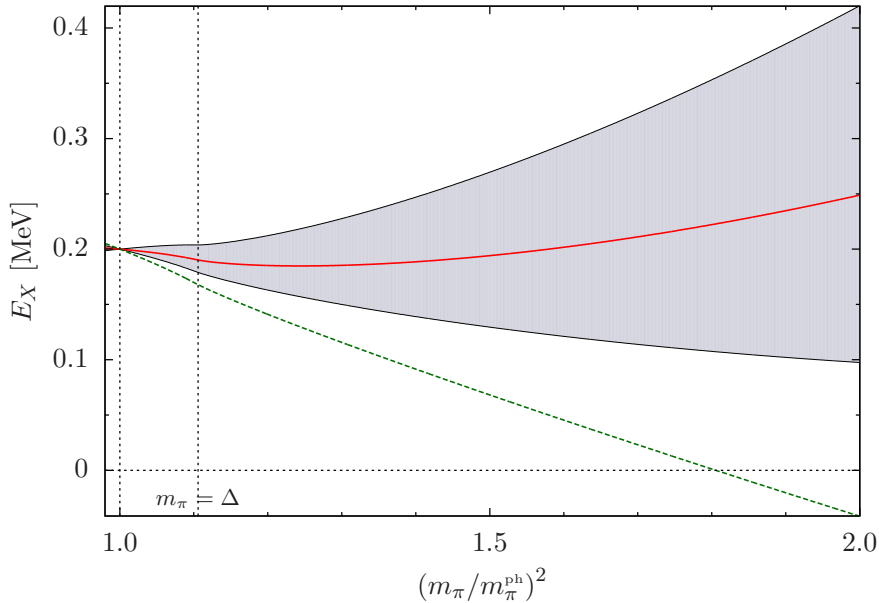}
	\end{center}
	\caption{Binding energy of the $X(3872)$. The solid, thick curve belongs to $d_2=0$ and $r_0=0$, i.e. considering the LO contact interaction and pion exchanges only. For $d_2=-1$ and $r_0=0.01/\tn{MeV}$ we acquire the lower bound and for $d_2=1$ and $r_0=0$ the upper bound for the binding energy. The dashed, thick curve corresponds to $d_2=-2$.}
	\label{fig:XB}
\end{figure}

In Fig.~\ref{fig:XB}, the quark mass dependence of the binding energy of the $X$ is shown. We plot against the squared pion mass, which is proportional to the light quark masses at leading-order in chiral perturbation theory. Since in XEFT power counting $\Lambda\gtrsim Q$, we use $\Lambda=50\tn{ MeV}$. As described in section \ref{sec:deltab}, we fix the binding energy at the physical value of the pion mass $m_\pi^\tn{ph}$ and use $E_X^\tn{ph}=0.2\tn{ MeV}$. The $m_\pi$-dependent and independent contact interactions at NLO can be tuned by modifying the parameters $d_2$ and $r_0$ defined in Eqs.~(\ref{eq:renord2}, \ref{eq:renorc2}), respectively. For $d_2=0$ and $r_0=0$, the $D$ mesons interact via the LO contact interaction and pion exchanges only, corresponding to the solid, thick curve in Fig.~\ref{fig:XB}. We see that for increasing pion mass the binding energy first moves towards the threshold with an inflection point at $m_\pi=\Delta$. Shortly after the inflection point, the sign of the slope reverses and the binding energy increases for increasing quark masses. 
Tuning the strengths of the NLO contact interactions via the parameters $d_2$ and $r_0$ can either imply that the slope of the binding energy of the $X$ enlarges or decreases. The lower bound for the binding energy is acquired for $d_2=-1$ and $r_0=0.01/\tn{MeV}$. For this scenario, the binding energy remains below the physical one for pion masses $\l(m_\pi^\tn{ph}\r)^2<m_\pi^2<2\l(m_\pi^\tn{ph}\r)^2$. On the other hand, assuming positive values for $d_2$ and small values for $r_0$, the binding energy of the $X$ steeply rises for pion masses beyond the inflection point. The upper bound belongs to $d_2=1$ and $r_0=0$. The dominating contribution to the shift of the binding energy at NLO is the quark mass dependent contact interaction. Considering, e.g., an unnaturally large and negative coupling constant $d_2$, it is possible that the bound state of the $X$ vanishes at higher quark masses. This is represented by the dashed, thick curve in Fig.~\ref{fig:XB} belonging to $d_2=-2$ and $r_0=0$. 

Before we consider the chiral extrapolation of the real part of the scattering length over the whole range of validity, let us take a closer look at its behavior around $m_\pi=\Delta$, i.e. where the pion mass is close to the hyperfine splitting of the $D^0$ and $D^{*0}$. At LO the scattering length reads
\begin{align}
	a^\tn{LO}&=\frac{1}{\gamma-\eta_0}=\frac{1}{\gamma-\sqrt{-2i\frac{M_{DD^*}g^2}{24\pi f^2}}\mu^{\frac{3}{2}}\theta\l(\Delta-m_\pi\r)}.
	\label{eq:sclLO}
\end{align}
For $m_\pi\rightarrow\Delta$, $\eta_0\rightarrow0$ and therefore $a^\tn{LO}\rightarrow1/\gamma$. The scattering length is continous but not differentiable at $m_\pi=\Delta$. We consider the derivative of the real part of the scattering length with respect to $m_\pi$ around $m_\pi=\Delta$ ($\mu=0$):
\begin{equation}
	\frac{\partial \operatorname{Re}\l[a^\tn{LO}\r]}{\partial m_\pi}=\begin{cases}
		-\frac{3}{2}\frac{m_\pi}{\gamma^2}\sqrt{\frac{M_{DD^*}g^2}{24\pi f^2}}\frac{1}{\sqrt{\mu}}+\mathcal{O}\l(\mu\r) &, m_\pi<\Delta,\\
		0											&, m_\pi>\Delta,
	\end{cases}
	\label{eq:delsclLO}
\end{equation}
i.e. the derivative is discontinuous at $m_\pi=\Delta$. It diverges to $-\infty$ for $m_\pi\nearrow\Delta$ and is zero for $m_\pi>\Delta$. This leads to a cusp effect for the scattering length at LO.

At NLO the cusp effect is smeared out due to the logarithmic term in $\mathcal{A}_0^\tn{(III)}$. Its contribution to the scattering length reads
\begin{equation}
	a^\tn{NLO}_\tn{(III)}=-\frac{M_{DD^*}g^2}{6\pi f^2}\mu^2\frac{1}{\l(i\mu-\eta_0\r)\l(\gamma-\eta_0\r)}.
	\label{eq:sclNLOiii}
\end{equation}
Using $\mu=i\l|\mu\r|$ for $m_\pi>\Delta$, it follows for the derivative of the real part with respect to $m_\pi$
\begin{equation}
	\frac{\partial \operatorname{Re}\l[a^\tn{NLO}_\tn{(III)}\r]}{\partial m_\pi}=\begin{cases}
		-6\frac{m_\pi}{\gamma}\sqrt{\frac{M_{DD^*}g^2}{24\pi f^2}}^3\frac{1}{\sqrt{\mu}}+\mathcal{O}\l(\mu^0\r) &, m_\pi<\Delta,\\
		-3\frac{m_\pi}{\gamma}\frac{M_{DD^*}g^2}{24\pi f^2}\frac{1}{\l|\mu\r|}			&, m_\pi>\Delta.
	\end{cases}
	\label{eq:delsclNLOiii}
\end{equation}
This implies that at NLO the derivative of the real part of the scattering length diverges to $-\infty$ for both limits, $m_\pi\nearrow\Delta$ and $m_\pi\searrow\Delta$.

\begin{figure}[ht]
	\begin{center}
		\includegraphics[width=0.7\textwidth]{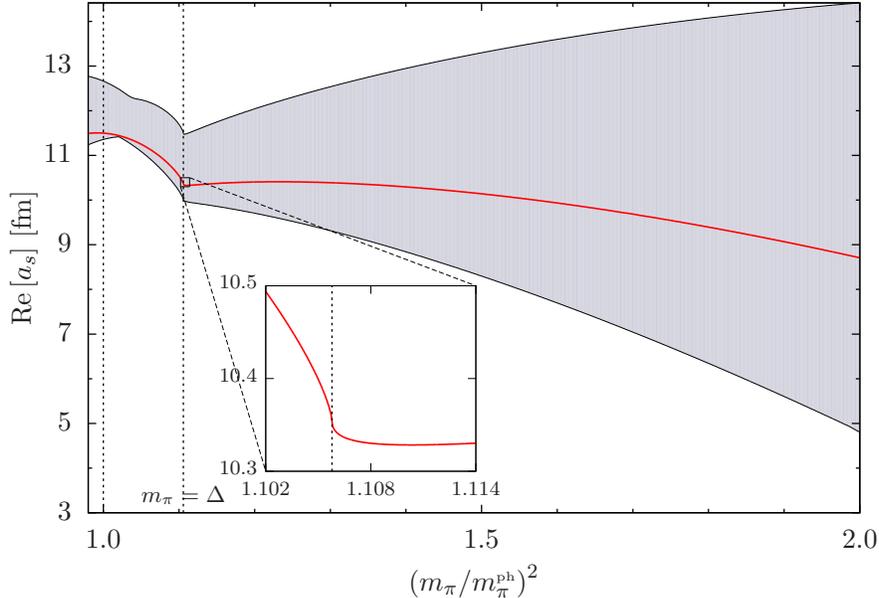}
	\end{center}
	\caption{The quark mass dependence of the real part of the $X(3872)$'s scattering length. The solid, thick curve belongs to $d_2=0$ and $r_0=0$. The bounds are acquired by varying $d_2$ and $r_0$ in their natural ranges $r_0\in \l[0,1/100\tn{MeV}\r]$
and $d_2\in \l[-1,1\r]$ and maximizing the width of the error band.}
	\label{fig:scl}
\end{figure}

The real part of the scattering length in dependence on the light quark masses is shown in Fig.~\ref{fig:scl}. We see the expected negative correlation between the scattering length and the binding energy, i.e. that the scattering length decreases for increasing binding energy and vice versa. 
Again the solid, thick curve belongs to the case where $d_2=0$ and $r_0=0$ and only the LO contact interaction and the pion exchanges are considered. 
The magnified area shows the smeared out cusp at $m_\pi=\Delta$ in more detail. 
The lower and upper bounds are obtained by varying $d_2$ and $r_0$ in the natural ranges
$r_0\in \l[0,1/100\tn{MeV}\r]$, $d_2\in \l[-1,1\r]$ and maximizing the width of the error band.
The scattering length is unnaturally large for all values of $m_\pi$. It is therefore expected, that XEFT power counting remains valid for pion masses beyond the physical one.

\section{Conclusion and Outlook}
\label{sec:conc}
In this work, we have investigated the light quark mass dependence of the $X(3872)$ to
NLO in XEFT where pions are included perturbatively.
We demonstrated that transition amplitudes containing dressed $D^{*0}$ propagators as subdiagrams 
exhibit infrared divergences and eliminated these divergences by using full $D^{*0}$ propagators.
Moreover, we have calculated $\bar{D}^0D^{*0}$ scattering in the $X$ channel and gave analytical
expressions for the binding energy of the $X(3872)$ and the $\bar{D}^0D^{*0}$ scattering length at NLO
in Eqs. \eqref{eq:EXNLO} and \eqref{eq:scl}.

At this order, the quark mass dependence of the $X$ is determined by the quark mass dependence of the 
$D^{(*)}$ masses, of the pion exchange interaction as well as a quark mass dependent contact
interaction that is required for consistent renormalization. In analogy to the $NN$ case \cite{Kaplan:1998tg},
our calculations demonstrate that it is essential to include such a term at NLO to obtain renormalization 
scale independent transition amplitudes. This invalidates the claim
of Wang and Wang \cite{Wang:2013kva} that the properties of the $X(3872)$ are determined 
by pion exchange alone. A similar conclusion was reached by Baru et al.~\cite{Baru:2013rta}
in a Faddeev-approach with a $D\bar{D}\pi$ contact interaction. 

Taking the quark mass dependence into account, there are two unknown constants $r_0$ and $d_2$ 
from contact interactions in the expressions for the binding energy and scattering length at NLO.
While $r_0$ could be determined by scattering data in the $X$ channel, the parameter $d_2$, which
governs the quark mass dependence of the NLO contact interaction, can only be determined on the lattice.
We constrained these coupling constants by dimensional analysis arguments and examined different scenarios
for their values. We found that it is most likely that the $X(3872)$ is bound for quark masses larger
than the physical one. However, for a unnaturally large and negative coupling constant $d_2$, a 
disappearance of the bound state for increasing quark masses is also possible. The qualitative behavior 
of the binding energy is in agreement with the results of Baru et al.~\cite{Baru:2013rta}.

Our predictions could be used to extrapolate lattice calculations of the $X(3872)$ at unphysical
quark masses to the physical ones. Based on a conservative estimate of higher order effects, our
results should be applicable in the region $0.98\l(m_\pi^\tn{ph}\r)^2
\lesssim m_\pi^2\lesssim2\l(m_\pi^\tn{ph}\r)^2$, where $m_\pi^\tn{ph}$ is the physical pion mass.
Our results suggest that it should be possible to find the $X$ in lattice simulations at 
quark masses in this region. Depending on the values of $d_2$ and $r_0$, it
can be more or less bound at larger quark masses. The first lattice results for the $X$ in full
 QCD were recently provided by Prelovsek and Leskovec \cite{Prelovsek:2013cra}. 
In a relatively small box with $L \approx 2$ fm, they found a candidate for the $X(3872)$ about $11\pm7\tn{ MeV}$ 
below the $\bar{D}^0D^{*0}$ threshold for a pion mass about twice the physical value and thus 
slightly beyond the range of applicability of our calculation.  If it is nevertheless
extrapolated to larger pion masses and finite volume effects are neglected, 
our result is consistent with the value of Prelovsek and Leskovec at the two sigma level.
If simulations for multiple smaller pion masses are carried out in the future, our 
calculations could be used to extrapolate the lattice results to the physical values.

In the future,  it would be interesting to extend our work to NNLO. At this order 
charged $D^{(*)}$ mesons will appear for the first time and lead to coupled channel effects 
if they are not integrated out of the theory. Furthermore, relativistic correction for the 
pions will appear and $m_\pi/m_D$ corrections in NLO diagrams have to be considered. Given
the small volumes currently available, it would also be useful to explicitly 
calculate finite volume effects in the binding energy and other observables within
XEFT.

\begin{acknowledgments}
We thank C. Hanhart and U.-G. Mei{\ss}ner for helpful discussions.
This research was supported in part by the DFG and the NSFC through funds provided to the Sino-German CRC 110, 
by the BMBF under grant 05P12PDFTE, by the NNSFC under grant 10935012, and by the
Helmholtz Association under contract HA216/EMMI.
\end{acknowledgments}

\numberwithin{equation}{section}
\appendix
\section{Calculation of the $D^{*0}$ self energy and $\mathcal{A}_0^\tn{(VI)}$}
\label{sec:sigmaaoVI}
In this section we show the explicit calculations for the diagrams depicted in Figs.~\ref{fig:dstbub} and \ref{fig:dstse}. We start with the bare $D^{*0}$ self energy diagram with spin indices $i$ and $j$
\begin{equation}
	i\hat{\Sigma}_{ij}=\l(\frac{\Lambda}{2}\r)^{4-D}\int \frac{d^Dq}{\l(2\pi\r)^D}\frac{-g^2}{2f^2}\frac{1}{2m_\pi}\frac{i}{-q_0-q^2/2m_D+i\epsilon}\frac{i\l(\boldsymbol{\varepsilon}_i\cdot\l(\mathbf{p+q}\r)\r)\l(\boldsymbol{\varepsilon}_j\cdot\l(\mathbf{p+q}\r)\r)}{p_0+q_0-(\mathbf{p+q})^2/2m_\pi+\delta+i\epsilon}.
	\label{eq:sigmadstar1}
\end{equation}
Performing the contour integration for $q_0$ we acquire
\begin{equation}
	i\hat{\Sigma}_{ij}=\frac{-ig^2}{2f^2}\frac{1}{2m_\pi}\l(\frac{\Lambda}{2}\r)^{4-D}\int \frac{d^{D-1}q}{\l(2\pi\r)^{D-1}}\frac{\l(\boldsymbol{\varepsilon}_i\cdot\l(\mathbf{p+q}\r)\r)\l(\boldsymbol{\varepsilon}_j\cdot\l(\mathbf{p+q}\r)\r)}{p_0-q^2/2m_D-(\mathbf{p+q})^2/2m_\pi+\delta+i\epsilon}.
	\label{eq:sigmadstar2}
\end{equation}
Using the rotational invariance of \eqref{eq:sigmadstar2} we can replace
\begin{equation}
	\boldsymbol{\varepsilon}_i\cdot\l(\mathbf{p+q}\r)\boldsymbol{\varepsilon}_j\cdot\l(\mathbf{p+q}\r)=\l(\mathbf{p+q}\r)_i\cdot\l(\mathbf{p+q}\r)_j\rightarrow\frac{\delta_{ij}}{D-1}\l(\mathbf{p+q}\r)^2
	\label{eq:replacepolarization}
\end{equation}
in the integral and obtain for the self energy diagram
\begin{equation}
	i\hat{\Sigma}_{ij}=\frac{ig^2}{2f^2}\frac{\delta_{ij}}{D-1}\l(\frac{\Lambda}{2}\r)^{4-D}\int \frac{d^{D-1}q}{\l(2\pi\r)^{D-1}}\frac{(\mathbf{p+q})^2}{2m_\pi p_0-q^2 m_\pi/m_D-(\mathbf{p+q})^2+2m_\pi\delta+i\epsilon}.
	\label{eq:sigmadstar3}
\end{equation}
The energy of the $D^{*0}$ meson is of order $p^2/2m_{D^*}$. As explained in section \ref{sec:ddstartrans}, diagrams including a pion with additional $m_\pi/m_D$ suppression are in the same order as the two-pion exchange graph which occurs at NNLO first and can be neglected. We further use that $2m_\pi\delta=\mu^2+\mathcal{O}\l(\delta/m_\pi\r)$. Similar to \eqref{eq:spindepamp} we utilize $i\hat{\Sigma}_{ij}=\delta_{ij}i\Sigma$. The self energy $i\Sigma$ is therefore approximately given as
\begin{align}
	i\Sigma\approx&-\frac{ig^2}{2f^2}\frac{1}{D-1}\l(\frac{\Lambda}{2}\r)^{4-D}\int \frac{d^{D-1}q}{\l(2\pi\r)^{D-1}}\frac{(\mathbf{p+q})^2}{(\mathbf{p+q})^2-\mu^2-i\epsilon}\notag\\
	=&-\frac{ig^2}{2f^2}\frac{1}{D-1}\l(\frac{\Lambda}{2}\r)^{4-D}\frac{1}{\l(4\pi\r)^{\l(D-1\r)/2}}\frac{D-1}{2}\Gamma\l[\frac{1-D}{2}\r]\l(-\mu^2-i\epsilon\r)^\frac{D-1}{2}.
	\label{eq:sigmadstar4}
\end{align}
In PDS renormalization scheme, we have to remove all poles in 3 and 4 dimensions. Taking the limit $D\rightarrow4$ we end up with
\begin{equation}
	i\Sigma=\frac{ig^2}{24\pi f^2}\l(i\mu^3+\Lambda\mu^2\r).
	\label{eq:sigmadstA}
\end{equation}
Using on-shell renormalization, the counter term has to remove the imaginary part of $i\Sigma$. This implies for the on-shell renormalized self energy
\begin{equation}
	i\Sigma^\tn{OS}=\begin{cases}
		-\frac{ig^2}{24\pi f^2}i\mu^3 &, m_\pi<\Delta,\\
		0 &, m_\pi\geq\Delta.
	\end{cases}
	\label{eq:deltamA}
\end{equation}
It follows for the transition amplitude $\mathcal{A}_0^\tn{(VI)}$ in Fig.~\ref{fig:dstbub}, utilizing the previous results
\begin{align}
	i\mathcal{A}_0^\tn{(VI)}=&i\mathcal{A}_{-1}\l(\frac{\Lambda}{2}\r)^{4-D}\int\frac{d^Dq}{\l(2\pi\r)^D}\l(\frac{i}{E+q_0-q^2/2m_{D^*}+i\epsilon}\r)^2i\Sigma^\tn{OS}\frac{i}{-q_0-q^2/2m_D}i\mathcal{A}_{-1}\notag\\
	=&i\mathcal{A}_{-1}^2\Sigma^\tn{OS}\l(\frac{\Lambda}{2}\r)^{4-D}\int\frac{d^{D-1}q}{\l(2\pi\r)^{D-1}}\l(\frac{2M_{DD^*}}{q^2-p^2-i\epsilon}\r)^2\notag\\
	=&i\mathcal{A}_{-1}^2\l(2M_{DD^*}\r)^2\Sigma^\tn{OS}\l(\frac{\Lambda}{2}\r)^{4-D}\frac{1}{\l(4\pi\r)^{\l(D-1\r)/2}}\Gamma\l[\frac{5-D}{2}\r]\l(-p^2-i\epsilon\r)^{\frac{D-5}{2}}.
	\label{eq:a0VIA}
\end{align}
This expression is ultraviolet finite in 3 and 4 dimensions. In the limit $D\rightarrow4$ we acquire the result given in Eq. \eqref{eq:a0VI}.

\bibliography{ref}


\end{document}